\documentclass{article}
\usepackage{spconf}

\usepackage{amsmath}
\usepackage{amsfonts}
\usepackage{amssymb,bm}
\usepackage{amsthm}
\usepackage{algorithm}
\usepackage{algpseudocode}
\usepackage{cite}
\usepackage{graphicx}
\usepackage{epstopdf}
\usepackage[caption=false,font=footnotesize]{subfig}
\usepackage{fixltx2e}
\usepackage{balance}
\usepackage{cases}

\title{A Rate-Splitting Approach To Robust Multiuser MISO Transmission}
\name{Hamdi Joudeh$^{\ast}$ and Bruno Clerckx$^{\ast \dagger}$}
\address{
$^{\ast}$ Department of Electrical and Electronic Engineering, Imperial College London, United Kingdom
\\
$^{\dagger}$ School of Electrical Engineering, Korea University, Seoul, Korea
\\
\fontsize{10}{10}\selectfont\ttfamily\upshape
Email: \{hamdi.joudeh10, b.clerckx\}@imperial.ac.uk
}
\begin{document}
\ninept
\maketitle
\begin{abstract}
For multiuser MISO systems with bounded uncertainties in the Channel State Information (CSI), we consider two classical robust design problems: maximizing the minimum rate subject to a transmit power constraint, and power minimization under a rate constraint.
Contrary to conventional strategies, we propose a Rate-Splitting (RS) strategy where each message is divided into two parts, a common part and a private part.
All common parts are packed into one \emph{super} common message encoded using a shared codebook and decoded by all users, while private parts are independently encoded and retrieved by their corresponding users.
We prove that RS-based designs achieve higher max-min Degrees of Freedom (DoF) compared to
conventional  designs (NoRS) for uncertainty regions that scale with SNR. For the special case of non-scaling
uncertainty regions, RS contrasts with NoRS and achieves a non-saturating max-min rate.
In the power minimization problem, RS is shown to combat the feasibility problem arising from multiuser interference in NoRS.
A robust design of precoders for RS is proposed, and performance gains over NoRS are demonstrated through simulations.
\end{abstract}
\section{Introduction}
\label{Section_Introduction}
\newcounter{Proposition_Counter} 
\newcounter{Theorem_Counter} 
\newcounter{Lemma_Counter} 
\newcounter{Remark_Counter} 
\newcounter{Assumption_Counter}
Consider a Multi-User (MU) Multiple Input Single Output (MISO) system where a Base Station (BS) equipped with $N_t$ antennas serves a set of single-antenna users $\mathcal{K} \triangleq \{1,\ldots,K\}$, with $K \leq N_{t}$.
Under the assumption of erroneous Channel State Information at the Transmitter (CSIT) with bounded uncertainty regions, we address two typical design problems: maximizing the minimum worst-case rate subject to a total transmit power constraint (rate problem), and minimizing the transmit power subject to a worst-case rate constraint (power problem).
Such problems, which are non-convex in general and semi-infinite, were addressed in literature using various approaches and approximations \cite{Vucic2009,Vucic2009a,Payaro2007,Shenouda2007,Shenouda2009}.
However, all existing works consider a conventional transmission scheme, i.e. each message is encoded into an independent data stream, then all streams are spatially multiplexed.
For the rate problem, such designs are known to yield a saturating performance at high SNRs where MU interference becomes dominant \cite{Jindal2006,Shenouda2007,Shenouda2009}.
Conversely, this creates a feasibility issue for the power problem, since rates beyond the saturation level cannot be achieved \cite{Shenouda2007,Vucic2009}.
In this work, we propose a Rate-Splitting (RS) strategy to combat these shortcomings.

Let $W_{\mathrm{t},1},\ldots,W_{\mathrm{t},K}$ be messages uniformly drawn from the sets $\mathcal{W}_{\mathrm{t},1},\ldots,\mathcal{W}_{\mathrm{t},K}$, and intended for receivers $1,\ldots,K$ respectively.
In the proposed scheme, each user message is split into a private part $W_{k} \in \mathcal{W}_{k}$ and a common part $W_{\mathrm{c},k} \in \mathcal{W}_{\mathrm{c},k}$,
where $\mathcal{W}_{k} \times \mathcal{W}_{\mathrm{c},k} = \mathcal{W}_{\mathrm{t},k}$.
A \emph{super} message (the common message) is composed by packing the common parts such that $W_{\mathrm{c}} = \{W_{\mathrm{c},1} ,\ldots,W_{\mathrm{c},K} \} \in \mathcal{W}_{\mathrm{c}}$,
where $\mathcal{W}_{\mathrm{c}} = \mathcal{W}_{\mathrm{c},1} \times \ldots \times \mathcal{W}_{\mathrm{c},K} $.
$W_{\mathrm{c}}$ is encoded using a common codebook shared by all users, while $W_{1},\ldots,W_{K}$ are encoded using private codebooks known by their corresponding users.
The resulting $K+1$ encoded symbol streams are linearly precoded and simultaneously transmitted.
At each receiver, the common message is decoded first by treating all private signals as noise. This is followed by decoding the private message after removing the common message via Successive Interference Cancellation (SIC). The original messages are delivered given that each receiver successfully decodes the common message and its private message.

A special case of the described RS scheme \cite{Yang2013,Hao2013}, where only one user message is split, was shown to boost the sum Degrees of Freedom (DoF) under CSIT errors that decay with increased SNR at a rate of $O(\mathrm{SNR}^{-\alpha})$ for some constant
$\alpha \in [0,1]$.
This strategy was leveraged to enhance the sum-rate performance under various CSIT assumptions \cite{Hao2015,Joudeh2015}.
RS was also shown to provide significant sum-rate gains in the large-scale array regime \cite{Dai2015a}.
Employing RS in a more general manner to achieve max-min fairness was first reported in our previous work \cite{Joudeh2015a}, where the average performance over the error distribution was considered.
This paper gives a more complete treatment of the max-min problem by deriving the asymptotic rate performance (in a DoF sense) of the optimally designed RS scheme. In addition, the approach is extended to tackle the inverse power optimization problem. We should also highlight that the worst-case optimization considered in this work poses an extra challenge in comparison to \cite{Joudeh2015a} due to the minimization embedded in each worst-case rate expression.
The rest of the paper is organized as follows. The system model is described in Section \ref{Section_System_Model}.
In Section \ref{Section_Problem_Statement}, the problem is formulated and the asymptotic performance is derived.
An optimized design of precoders for the RS strategy is proposed in Section \ref{Section_Optimization}.
Simulation results and analysis are presented in Section \ref{Section_Simulation_Results}, and Section \ref{Section_Conclusion} concludes
the paper.
\section{System Model}
\label{Section_System_Model}
For the system introduced in the previous section, consider a transmission taking place over a block of channel uses where the channel remains fixed (quasi-static).
The signal received at the $k$th user in a given channel use writes as $y_{k} =  \mathbf{h}_{k}^{H}\mathbf{x}+n_{k}$,
where $\mathbf{h}_{k} \in \mathbb{C}^{N_{t}}$ is the channel vector from the BS to the $k$th user,  $\mathbf{x} \in \mathbb{C}^{N_{t}}$ is the transmit signal, and $n_{k} \thicksim \mathcal{CN} ( 0 , \sigma^{2}_{\mathrm{n},{k}} )$ is the Additive White Gaussian Noise (AWGN) at the receiver.
The transmit signal is subject to an average power constraint $\mathrm{E}\{\mathbf{x}^{H}\mathbf{x}\} \leq P_{\mathrm{t}}$.
Without loss of generality, we assume equal noise variances across users, i.e. $\sigma_{\mathrm{n},{k}}^{2}=\sigma_{\mathrm{n}}^{2}$, and channel entries with normalized average gains.
Therefore, the long term SNR is defined as $\mathrm{SNR} \triangleq P_{\mathrm{t}} / \sigma_{\mathrm{n}}^{2}$.
Moreover, $\sigma_{\mathrm{n}}^{2}$ is non-zero and remains fixed. Hence, $P_{\mathrm{t}}  \rightarrow \infty$ implies $\mathrm{SNR} \rightarrow \infty$.
$W_{\mathrm{c}},W_{1},\ldots,W_{K}$ are encoded into the independent data streams $s_{\mathrm{c}},s_{1},\ldots,s_{K}$ respectively.
For a given channel use, symbols are grouped as $\mathbf{s} \triangleq [s_{\mathrm{c}},s_{1},\ldots,s_{K}]^{T} \in \mathbb{C}^{K+1}$, where
$\mathrm{E}\{\mathbf{s}\mathbf{s}^{H}\} = \mathbf{I}$.
The symbols are mapped to the BS antennas through the precoding matrix $\mathbf{P} \triangleq [\mathbf{p}_{\mathrm{c}},\mathbf{p}_{1},\ldots,\mathbf{p}_{K}]$, yielding the transmit signal
$\mathbf{x} = \mathbf{P}\mathbf{s}$,
where the common stream is superimposed on top of the private streams. The power constraint is rewritten as $\mathrm{tr}\big( \mathbf{P}\mathbf{P}^{H} \big) \leq P_{\mathrm{t}}$.
The $k$th average receive power is expressed as
\begin{equation}
\label{Eq_T_c_k}
T_{\mathrm{c},k} = \overbrace{|\mathbf{h}_{k}^{H}\mathbf{p}_{\mathrm{c}}|^{2}}^{S_{\mathrm{c},k}} + \underbrace{ \overbrace{|\mathbf{h}_{k}^{H}\mathbf{p}_{k}|^{2}}^{S_{k}} + \overbrace{\sum_{i\neq k} |\mathbf{h}_{k}^{H}\mathbf{p}_{i}|^{2} + \sigma_{\mathrm{n}}^{2}}^{I_{k}}}_{I_{\mathrm{c},k} = T_{k}}.
\end{equation}
The common message is first decoded by treating interference from all private signals as noise.
The $k$th user's estimate of $s_{\mathrm{c}}$ is obtained as $\widehat{s}_{\mathrm{c},k}=g_{\mathrm{c},k} y_{k}$,  where $g_{\mathrm{c},k} $ is a scalar equalizer.
Given that the common message is successfully decoded, SIC is used to remove the common signal from $y_{k}$ before decoding the private message.
The estimate of $s_{k}$ writes as $\widehat{s}_{k}=g_{k}(y_{k}-\mathbf{h}_{k}^{H}\mathbf{p}_{\mathrm{c}}s_{\mathrm{c},k})$,
where $g_{k}$ is the corresponding equalizer.
At the output of the $k$th receiver, the common and private MSEs, defined as $\varepsilon_{\mathrm{c},k} \triangleq \mathrm{E}\{|\widehat{s}_{\mathrm{c},k} - s_{\mathrm{c}}|^{2}\}$ and $\varepsilon_{k} \triangleq \mathrm{E}\{|\widehat{s}_{k} - s_{k}|^{2}\}$ respectively, write as:
\begin{subequations}
\label{Eq_MSE}
\begin{align}
  \label{Eq_MSE_c_k}
  \varepsilon_{\mathrm{c},k} & = |g_{\mathrm{c},k}|^{2} T_{\mathrm{c},k} -2\Re \big\{g_{\mathrm{c},k}\mathbf{h}_{k}^{H}\mathbf{p}_{\mathrm{c}}\big\}+1 \\
  \label{Eq_MSE_k}
  \varepsilon_{k} & =  |g_{k}|^{2} T_{k}-2\Re \big\{g_{k}\mathbf{h}_{k}^{H}\mathbf{p}_{k}\big\}+1
\end{align}
\end{subequations}
where $T_{k}$ is defined in \eqref{Eq_T_c_k}.
We assume perfect CSI at the Receivers (CSIR) obtained through common and dedicated downlink training.
This assumption is justified by noting that the effects of CSIR estimation errors are at the same power level of additive noise, which is overwhelmed by the influence of CSIT uncertainty \cite{Caire2010}.
Users employ optimum  $(g_{\mathrm{c},k},g_{k})$, i.e. the well-known Minimum MSE (MMSE) equalizers given as:
$g_{\mathrm{c},k}^{\mathrm{MMSE}} = \mathbf{p}_{\mathrm{c}}^{H}\mathbf{h}_{k} T_{\mathrm{c},k}^{-1}$ and
$g_{k}^{\mathrm{MMSE}} = \mathbf{p}_{k}^{H}\mathbf{h}_{k}T_{k}^{-1}$,
from which the  MMSEs are obtained as:
$\varepsilon_{\mathrm{c},k}^{\mathrm{MMSE}}  =  T_{\mathrm{c},k}^{-1} I_{\mathrm{c},k}$
and
$\varepsilon_{k}^{\mathrm{MMSE}} = T_{k}^{-1}I_{k}$.
The $k$th SINRs write as: $\gamma_{\mathrm{c},k}  = S_{\mathrm{c},k}I_{\mathrm{c},k}^{-1} = (1 - \varepsilon_{\mathrm{c},k}^{\mathrm{MMSE}} )/\varepsilon_{\mathrm{c},k}^{\mathrm{MMSE}}$
and $\gamma_{k}  = S_{k}I_{k}^{-1} = (1 - \varepsilon_{k}^{\mathrm{MMSE}} )/\varepsilon_{k}^{\mathrm{MMSE}}$.
Under Gaussian signalling, the $k$th maximum achievable rates are given as:
$R_{\mathrm{c},k} =  \log_{2}(1+\gamma_{\mathrm{c},k})=-\log_{2}(\varepsilon_{\mathrm{c},k}^{\mathrm{MMSE}})$
and
$R_{k} = \log_{2}(1+\gamma_{k})=-\log_{2}(\varepsilon_{k}^{\mathrm{MMSE}})$.
To ensure that all users decode $W_{\mathrm{c}}$, it is transmitted in a multicast fashion at the common rate $R_{\mathrm{c}} \triangleq \min_{j}\{R_{\mathrm{c},j}\}_{j=1}^{K}$.
%
\section{Problem Statement and Asymptotic Performance}
\label{Section_Problem_Statement}
For each channel vector $\mathbf{h}_{k}$, the BS obtains an erroneous estimate $\widehat{\mathbf{h}}_{k}$,
from which the error unknown to the BS is defined as  $\widetilde{\mathbf{h}}_{k} \triangleq \mathbf{h}_{k} -  \widehat{\mathbf{h}}_{k}$.
As far as the BS is aware, $\widetilde{\mathbf{h}}_{k}$ is bounded by an origin-centered sphere with radius $\delta_{k}$.
Hence, $\mathbf{h}_{k}$ is confined within the uncertainty region
$\mathbb{H}_{k} \triangleq \left\{ \mathbf{h}_{k} \mid  \mathbf{h}_{k} = \widehat{\mathbf{h}}_{k} +   \widetilde{\mathbf{h}}_{k},
\|\widetilde{\mathbf{h}}_{k}\| \leq \delta_{k}  \right\}$.
This is relevant in limited feedback systems as quantization errors are bounded.
In scenarios where the number of feedback bits is made to scale with increased SNR to provide improved CSIT quality \cite{Jindal2006,Hao2015}, the uncertainty region shrinks such that
$\delta_{k}^{2} = O(P_{\mathrm{t}}^{-\alpha})$, where  $\alpha \in [0,\infty)$ is a constant exponent that quantifies the CSIT quality as SNR grows large (assumed to be the same across users), i.e.
$\alpha \triangleq \lim_{P_{\mathrm{t}}\rightarrow\infty} -\frac{\log(\delta_{k}^{2})}{\log(P_{\mathrm{t}})}$.
$\alpha =0$ represents a constant (or slowly scaling) number of feedback bits, yielding non-decaying CSIT errors.
On the other hand,
$\alpha = \infty$ corresponds to perfect CSIT resulting from an infinitely high number of feedback bits.
In the following, the exponents are truncated such that $\alpha \in [0,1]$ which is customary in DoF analysis as $\alpha = 1$ corresponds to perfect CSIT in the DoF sense \cite{Jindal2006,Yang2013}.
%

%
%
Due to the CSIT uncertainty, the actual rates cannot be considered as design metrics at the BS.
From the BS's point of view, achievable rates also lie in bounded uncertainty regions.
We consider a robust design where precoders are optimized w.r.t the worst-case achievable rates defined for the $k$th user  as
$\bar{R}_{\mathrm{c},k} \triangleq {\min}_{\mathbf{h}_{k} \in \mathbb{H}_{k}}
R_{\mathrm{c},k}(\mathbf{h}_{k})$
and
$\bar{R}_{k} \triangleq {\min}_{\mathbf{h}_{k} \in \mathbb{H}_{k}}
R_{k}(\mathbf{h}_{k})$,
where the dependencies of the rates on $\mathbf{h}_{k}$ are highlighted. The worst-case achievable common rate is defined as
%
$\bar{R}_{\mathrm{c}} \triangleq \min_{j}\{\bar{R}_{\mathrm{c},j}\}_{j=1}^{K}$.
%
Transmitting $W_{\mathrm{c}},W_{1},\ldots,W_{K}$ at rates $\bar{R}_{\mathrm{c}},\bar{R}_{1},\ldots,\bar{R}_{K}$ respectively, guarantees successful decoding at the receivers for all admissible channels within the uncertainty regions.
Following the RS structure in Section \ref{Section_Introduction}, the $k$th user's portion of the common rate is denoted by $\bar{C}_{k}$ where $\sum_{k=1}^{K}\bar{C}_{k} = \bar{R}_{\mathrm{c}}$.
Therefore, the $k$th user's worst-case total achievable rate is given as $\bar{R}_{\mathrm{t},k} =\bar{R}_{k} + \bar{C}_{k}$, corresponding to the rate at which the original message $W_{\mathrm{t},k}$ is transmitted.
\subsection{Rate Optimization Problem}
Using the RS strategy, the robust rate optimization problem which achieves max-min fairness is posed as
%
\vspace{-0.05 in}
\begin{equation}
\label{Eq_Problem_R_RS}
\mathcal{R}_{\mathrm{RS}}(P_{\mathrm{t}}):
\begin{cases}
       \underset{ \bar{R}_{\mathrm{t}},\bar{R}_{\mathrm{c}},\bar{\mathbf{c}},\mathbf{P} }{\max} & \bar{R}_{\mathrm{t}} \\
       \text{s.t.}  & \bar{R}_{k} + \bar{C}_{k}  \geq  \bar{R}_{\mathrm{t}},\; \forall k \in  \mathcal{K} \\
                    & \bar{R}_{\mathrm{c},k} \geq \bar{R}_{\mathrm{c}}, \; \forall k\in\mathcal{K} \\
                    & \sum_{k=1}^{K} \bar{C}_{k} =  \bar{R}_{\mathrm{c}} \\
                    & \bar{C}_{k} \geq 0, \; \forall k \in \mathcal{K}\\
                    & \mathrm{tr}\big(\mathbf{P}\mathbf{P}^{H}\big) \leq P_{\mathrm{t}}.
\end{cases}
\end{equation}
where $\bar{R}_{\mathrm{t}}$ is an auxiliary variable, and $\bar{\mathbf{c}} \triangleq [\bar{C}_{1},\ldots,\bar{C}_{K}]^{T}$.
Pointwise minimizations in $\bar{R}_{\mathrm{t}}$ and $\bar{R}_{\mathrm{c}}$ are replaced with inequality constraints in \eqref{Eq_Problem_R_RS}, where equality holds at least for one user at optimality.
The constraints $\bar{C}_{k} \geq 0$ guarantee non-negative splitting.
In contrast, the NoRS version of the problem is formulated as
\vspace{-0.05 in}
\begin{equation}
\label{Eq_Problem_R_NoRS}
\mathcal{R}(P_{\mathrm{t}}):
\begin{cases}
       \underset{\bar{R},\mathbf{P}_{\mathrm{p}}}{\max} & \bar{R} \\
       \text{s.t.}  & \bar{R}_{k} \geq \bar{R}, \;  \forall k\in\mathcal{K} \\
                    & \mathrm{tr}\big(\mathbf{P}_{\mathrm{p}}\mathbf{P}_{\mathrm{p}}^{H}\big) \leq P_{\mathrm{t}}.
\end{cases}
\end{equation}
where $\bar{R}$ is the rate auxiliary variable, and $\mathbf{P}_{\mathrm{p}}\triangleq [\mathbf{p}_{1},\ldots,\mathbf{p}_{K}]$.
Solving \eqref{Eq_Problem_R_NoRS} is equivalent to solving \eqref{Eq_Problem_R_RS} over a restricted domain characterized by setting $\bar{\mathbf{c}} = \mathbf{0}$, which in turn forces $\bar{R}_{\mathrm{c}}$ and
$\|\mathbf{p}_{\mathrm{c}}\|^{2}$ to zeros at optimality. As a result, we have $\mathcal{R}_{\mathrm{RS}}(P_{\mathrm{t}}) \geq \mathcal{R}(P_{\mathrm{t}})$.
Next, we look at the optimum asymptotic performance of the two schemes.
%

%
To analyse the performance as SNR increases, we define a precoding scheme for \eqref{Eq_Problem_R_RS} as a family of feasible precoders with
one precoder for each SNR level,
i.e. $\big\{ \mathbf{P}(P_{\mathrm{t}}) \big\}_{P_{\mathrm{t}}}$.
The associated powers allocated to the precoding vectors are assumed to scale with $P_{\mathrm{t}}$ as $\| \mathbf{p}_{\mathrm{c}} \|^{2} = O(P_{\mathrm{t}}^{a_{\mathrm{c}}})$ and $\| \mathbf{p}_{k} \|^{2} = O(P_{\mathrm{t}}^{a_{k}})$, where
$a_{\mathrm{c}},a_{k} \in [0,1]$ are some scaling exponents.
Under a given precoding scheme, the $k$th worst-case common and private DoF are defined as
\vspace{-0.03 in}
\begin{equation}
\label{Eq_Wors_Case_DoF}
\bar{d}_{\mathrm{c}}
\triangleq \lim_{P_{\mathrm{t}} \rightarrow \infty} \frac{\bar{R}_{\mathrm{c}}(P_{\mathrm{t}})}{\log_{2}(P_{\mathrm{t}})}
\quad \text{and} \quad
\bar{d}_{k}
\triangleq \lim_{P_{\mathrm{t}} \rightarrow \infty} \frac{\bar{R}_{k}(P_{\mathrm{t}})}{\log_{2}(P_{\mathrm{t}})}
\end{equation}
where the dependencies on the power level are highlighted in \eqref{Eq_Wors_Case_DoF}.
The $k$th user's split of $\bar{d}_{\mathrm{c}}$ is defined as
$\bar{c}_{k} \triangleq \lim_{P_{\mathrm{t}} \rightarrow \infty} \frac{\bar{C}_{k}(P_{\mathrm{t}})}{\log_{2}(P_{\mathrm{t}})}$,
where $\sum_{k=1}^{K} \bar{c}_{k} = \bar{d}_{\mathrm{c}}$.
All definitions extend to the NoRS case where the common part is discarded, and a precoding scheme for \eqref{Eq_Problem_R_NoRS} is denoted by $\big\{ \mathbf{P}_{\mathrm{p}}(P_{\mathrm{t}}) \big\}_{P_{\mathrm{t}}}$.
%
\newtheorem{Theorem_RS_NoRS_Max_Min_DoF}[Theorem_Counter]{Theorem}
\begin{Theorem_RS_NoRS_Max_Min_DoF}\label{Theorem_RS_NoRS_Max_Min_DoF}
\textnormal{
The NoRS problem yields an optimum max-min DoF
\begin{equation}
\label{Eq_Max_Min_DoF_NoRS}
\bar{d}^{\ast} \triangleq \lim_{P_{\mathrm{t}} \rightarrow \infty} \frac{\mathcal{R}(P_{\mathrm{t}})}{\log(P_{\mathrm{t}})} = \alpha
\end{equation}
while the optimum max-min DoF for the RS problem is given by
\begin{equation}
\label{Eq_Max_Min_DoF_RS}
\bar{d}^{\ast}_{\mathrm{RS}} \triangleq \lim_{P_{\mathrm{t}} \rightarrow \infty} \frac{\mathcal{R}_{\mathrm{RS}}(P_{\mathrm{t}})}{\log(P_{\mathrm{t}})} =  \frac{1+ (K-1) \alpha}{K}.
\end{equation}
}
\end{Theorem_RS_NoRS_Max_Min_DoF}
\begin{proof}
Let $\{a_{k}^{\ast}\}_{k=1}^{K}$ be the power exponents of \eqref{Eq_Problem_R_NoRS}'s optimum precoding scheme. We initially assume that the optimum powers satisfy $a_{1}^{\ast},\ldots,a_{K}^{\ast} = a^{\ast}$. We have $\bar{d}_{k}^{\ast} \leq \min{\{\alpha,a^{\ast}\}}$.
This is shown through  upper-bounding the worst-case SINR, i.e. $\bar{\gamma}_{k} \triangleq \min_{\mathbf{h}_{k} \in \mathbb{H}_{k}} \gamma_{k}(\mathbf{h}_{k})$,  by selecting $\mathbf{h}_{k} \in \mathbb{H}_{k}$ such that the $l$th ($l \in \mathcal{K} \setminus k$) interference is maximized, and discarding some interference terms \cite{Tajer2011}.
Since the point-wise minimum is upper-bounded by any element in the set, and $\bar{d}^{\ast} \leq \bar{d}_{k}^{\ast}$ for all $k\in \mathcal{K}$,
we have $\bar{d}^{\ast} \leq \alpha$.
Next, assume that \eqref{Eq_Problem_R_RS}'s optimum precoding scheme has power exponents $a_{\mathrm{c}}^{\ast}$ and $a^{\ast}$ for the common and private precoders respectively.
Using $\bar{R}_{\mathrm{c}} + \bar{R}_{k} \leq \bar{R}_{\mathrm{c},k} + \bar{R}_{k}$, and the previous bounding techniques, we obtain $\bar{d}_{\mathrm{c}}^{\ast} + \bar{d}_{k}^{\ast} \leq \min\{1+\alpha - a^{\ast} , 1\}$.
Since the max-min DoF is upper-bounded by the average user DoF, we write
$\bar{d}_{\mathrm{RS}}^{\ast} \leq \frac{\bar{d}_{\mathrm{c}}^{\ast} + \sum_{k=1}^{K}\bar{d}_{k}^{\ast} }{K} \leq
\frac{1 + (K-1)\alpha }{K}$,
where the right-most inequality is obtained from $\bar{d}_{\mathrm{c}}^{\ast} + \bar{d}_{1}^{\ast} \leq 1$ and $\bar{d}_{k}^{\ast} \leq \alpha$.

The upper-bounds are achieved through feasible precoders as follows.
Best-effort Zero-Forcing (ZF) obtained using the available channel estimate with powers scaling as $O(P_{\mathrm{t}}^{\alpha})$ is used for private precoders, achieving $\bar{d}_{1},\ldots,\bar{d}_{K} = \alpha$. For RS, this is superimposed by a random common precoder with power that scales as $O(P_{\mathrm{t}})$, achieving $\bar{d}_{\mathrm{c}} = 1-\alpha$ which is split equally among users.
Relaxing the assumption that $a_{1}^{\ast},\ldots,a_{K}^{\ast}$ are equal yields the same result using slightly more involved upper-bounding steps.
\end{proof}
%
%
%
It should be noted that although the ZF precoders proposed in the proof are optimum in a DoF sense, they may be far from optimum from a worst-case rate perspective at finite SNRs. This motivates the need for robustly designed precoders.
It is evident that  $\bar{d}_{\mathrm{RS}}^{\ast} \geq \bar{d}^{\ast} $ holds for all $ \alpha \in [0,1]$, and strictly holds for $\alpha \in [0,1)$.
$\bar{d}^{\ast}_{\mathrm{RS}}$ is lower-bounded by $1/K$. Hence, an optimally designed RS scheme is expected to achieve an ever-growing max-min rate.
\subsection{Power Optimization Problem}
\label{Subsection_power_problem}
The  power problem with a rate constraint $\bar{R}_{\mathrm{t}}$ writes as
%
\vspace{-0.1 in}
\begin{equation}
\label{Eq_Problem_P_RS}
\mathcal{P}_{\mathrm{RS}}(\bar{R}_{\mathrm{t}}):
\begin{cases}
       \underset{\bar{R}_{\mathrm{c}}, \bar{\mathbf{c}} ,\mathbf{P} }{\min} & \mathrm{tr}\big(\mathbf{P}\mathbf{P}^{H}\big) \\
       \text{s.t.}  & \bar{R}_{k} + \bar{C}_{k}  \geq  \bar{R}_{\mathrm{t}},\; \forall k \in  \mathcal{K} \\
                    & \bar{R}_{\mathrm{c},k} \geq \bar{R}_{\mathrm{c}}, \; \forall k\in\mathcal{K} \\
                    & \sum_{k=1}^{K} \bar{C}_{k} =  \bar{R}_{\mathrm{c}} \\
                    & \bar{C}_{k} \geq 0, \; \forall k \in \mathcal{K}.
\end{cases}
\end{equation}
On the other hand, the NoRS counterpart is formulated as
%
\vspace{-0.1 in}
\begin{equation}
\label{Eq_Problem_P_NoRS}
\mathcal{P}(\bar{R}):
\begin{cases}
       \underset{\mathbf{P}_{\mathrm{p}} }{\min} & \mathrm{tr}\big(\mathbf{P}_{\mathrm{p}}\mathbf{P}_{\mathrm{p}}^{H}\big) \\
       \text{s.t.}  & \bar{R}_{k} \geq \bar{R}, \; \forall k\in\mathcal{K}.
\end{cases}
\end{equation}
As \eqref{Eq_Problem_P_NoRS} is a restricted version of \eqref{Eq_Problem_P_RS}, we have
$\mathcal{P}_{\mathrm{RS}}(\bar{R}_{\mathrm{t}}) \leq \mathcal{P}(\bar{R}_{\mathrm{t}})$.
We consider non-scaling CSIT with $\delta_{1}^{2},\ldots,\delta_{K}^{2} = O(1)$, i.e. $\alpha = 0$.
This is relevant to \eqref{Eq_Problem_P_RS} and \eqref{Eq_Problem_P_NoRS} where the CSIT quality does not change with the transmit power variation during the design procedure, as channel estimation and feedback is carried out prior to the precoder design.
The rate and the power problems are monotonically non-decreasing in their arguments, and are related such that $\mathcal{R}\big( \mathcal{P}(\bar{R}) \big) = \bar{R}$ and
$\mathcal{R}_{\mathrm{RS}}\big( \mathcal{P}_{\mathrm{RS}}(\bar{R}_{\mathrm{t}}) \big) = \bar{R}_{\mathrm{t}}$, which can be demonstrated using the same steps in \cite{Wiesel2006,Sidiropoulos2006}.
Combining this with Theorem \ref{Theorem_RS_NoRS_Max_Min_DoF}, it follows that under non-scaling CSIT errors, $\mathcal{R}\big( P_{\mathrm{t}} \big)$ converges to a finite maximum value as $P_{\mathrm{t}} \rightarrow \infty$, which is the maximum feasible rate for $\mathcal{P}(\bar{R})$.
On the other hand, $\mathcal{R}_{\mathrm{RS}}( P_{\mathrm{t}} )$ does not converge. Hence, any finite rate is feasible for $\mathcal{P}_{\mathrm{RS}}(\bar{R}_{\mathrm{t}})$.
\section{Robust Optimization}
\label{Section_Optimization}
Problems \eqref{Eq_Problem_R_RS} and \eqref{Eq_Problem_P_RS} are semi-infinite and appear to be intractable in their current forms. Even
finite instances of the problems seem to be intractable due to the non-convex coupled sum-rate expressions embedded in each user's total rate.
Therefore, we employ the Rate-Weighted MSE (WMSE) relationship which is particularly suitable for problems featuring sum-rate expressions \cite{Christensen2008,Shi2011,Razaviyayn2013b}.
The $k$th user's augmented WMSEs (referred to as WMSEs for brevity) are defined as:
$\xi_{\mathrm{c},k} \triangleq u_{\mathrm{c},k} \varepsilon_{\mathrm{c},k}  -  \log_{2} (  u_{\mathrm{c},k} )$ and
$\xi_{k} \triangleq u_{k} \varepsilon_{k}  -  \log_{2}  (  u_{k} )$,
with $u_{\mathrm{c},k}$ and $u_{k}$ as the corresponding weights.
Optimizing over the equalizers and weights, the Rate-WMSE relationship writes as:
$\xi_{\mathrm{c},k}^{\mathrm{MMSE}} \triangleq {\min}_{u_{\mathrm{c},k}, g_{\mathrm{c},k}} \xi_{\mathrm{c},k} = 1-R_{\mathrm{c},k}$
and
$\xi_{k}^{\mathrm{MMSE}} \triangleq {\min}_{u_{k}, g_{k}} \xi_{k}= 1-R_{k}$,
where the optimum equalizers and weights are given by:
$g_{\mathrm{c},k}^{\ast}  = g_{\mathrm{c},k}^{\mathrm{MMSE}}$,
$g_{k}^{\ast} = g_{k}^{\mathrm{MMSE}}$,
$u_{\mathrm{c},k}^{\ast} = \big( \varepsilon_{\mathrm{c},k}^{\mathrm{MMSE}} \big)^{-1}$,
and
$u_{k}^{\ast} = \big( \varepsilon_{k}^{\mathrm{MMSE}} \big)^{-1}$ \cite{Joudeh2015,Joudeh2015a}.
%
From this relationship, the worst-case rates are equivalently written as
\vspace{-0.03 in}
\begin{subequations}
\label{Eq_wc_Rates_WMSEs}
\begin{align}
\bar{R}_{\mathrm{c},k} & = 1 - \max_{\mathbf{h}_{k} \in \mathbb{H}_{k}} \min_{u_{\mathrm{c},k}, g_{\mathrm{c},k}}
\xi_{\mathrm{c},k}\big( \mathbf{h}_{k}, g_{\mathrm{c},k}, u_{\mathrm{c},k} \big) \\
\bar{R}_{k} & = 1 - \max_{\mathbf{h}_{k} \in \mathbb{H}_{k}} \min_{u_{k}, g_{k}}
\xi_{k}\big( \mathbf{h}_{k}, g_{k}, u_{k} \big).
\end{align}
\end{subequations}
Equivalent WMSE problems are obtained by substituting \eqref{Eq_wc_Rates_WMSEs} into
\eqref{Eq_Problem_R_RS} and \eqref{Eq_Problem_P_RS}, where the domains are extended to include the equalizers and weights as optimization variables.
Such problems have an interesting block-wise convex structure which can be exploited using the Alternating Optimization (AO) principle.
However, the new problems have infinitely many optimization variables and constraints due to the dependencies of the optimum equalizers and weights on perfect CSI.
We resort to the conservative approximation in \cite{Tajer2011}
by swapping the minimization and maximization in \eqref{Eq_wc_Rates_WMSEs}. Equalizers and weights loose their dependencies on perfect CSI and we obtain
\vspace{-0.03 in}
\begin{subequations}
\label{Eq_wc_Rates_WMSEs_LB}
\begin{align}
\label{Eq_wc_Rates_WMSEs_c_LB}
\widehat{R}_{\mathrm{c},k} & = 1 -  \min_{\widehat{u}_{\mathrm{c},k}, \widehat{g}_{\mathrm{c},k}} \max_{\mathbf{h}_{k} \in \mathbb{H}_{k}}
\xi_{\mathrm{c},k}\big( \mathbf{h}_{k}, \widehat{g}_{\mathrm{c},k}, \widehat{u}_{\mathrm{c},k} \big) \\
\widehat{R}_{k} & = 1 -  \min_{\widehat{u}_{k}, \widehat{g}_{k}} \max_{\mathbf{h}_{k} \in \mathbb{H}_{k}}
\xi_{k}\big( \mathbf{h}_{k}, \widehat{g}_{k}, \widehat{u}_{k} \big)
\end{align}
\end{subequations}
where $\widehat{R}_{\mathrm{c},k} \leq \bar{R}_{\mathrm{c},k}$ and
$\widehat{R}_{k} \leq \bar{R}_{k}$ are lower-bounds on the worst-case rates (see footnote 1 in \cite[Section IV.B.2]{Tajer2011}), and $(\widehat{g}_{\mathrm{c},k},\widehat{g}_{k})$ and $(\widehat{u}_{\mathrm{c},k},\widehat{u}_{k})$ are the abstracted equalizers and weights which are applied to all channels in the uncertainty sets.
Plugging \eqref{Eq_wc_Rates_WMSEs_LB}  into the rate problem yields the conservative WMSE counterpart
%
%
\vspace{-0.04 in}
\begin{equation}
\label{Eq_Problem_WMSE_R_RS_LB}
\widehat{\mathcal{R}}_{\mathrm{RS}}(P_{\mathrm{t}}) \! : \!
\begin{cases}
       \underset{ \widehat{R}_{\mathrm{t}},\widehat{R}_{\mathrm{c}},\widehat{\mathbf{c}},\mathbf{P},\widehat{\mathbf{g}},\widehat{\mathbf{u}}}
       {\max}
       \ \ \widehat{R}_{\mathrm{t}} \\
       \text{s.t.} \\
       1 \! - \! \xi_{k}\big( \mathbf{h}_{k}, \widehat{g}_{k}, \widehat{u}_{k} \big) \! + \! \widehat{C}_{k}  \geq  \widehat{R}_{\mathrm{t}}, \forall \mathbf{h}_{k} \in \mathbb{H}_{k}, k \in  \mathcal{K} \\
       1 - \xi_{\mathrm{c},k}\big( \mathbf{h}_{k}, \widehat{g}_{\mathrm{c},k}, \widehat{u}_{\mathrm{c},k} \big)  \geq \widehat{R}_{\mathrm{c}}, \forall \mathbf{h}_{k} \in \mathbb{H}_{k},
                    k \in  \mathcal{K} \\
                   \quad \sum_{k=1}^{K} \widehat{C}_{k} =  \widehat{R}_{\mathrm{c}} \\
                   \quad \widehat{C}_{k} \geq 0, \; \forall k \in \mathcal{K}\\
                   \quad \mathrm{tr}\big(\mathbf{P}\mathbf{P}^{H}\big) \leq P_{\mathrm{t}}
\end{cases}
\end{equation}
where $\widehat{\mathbf{g}} \triangleq \{\widehat{g}_{\mathrm{c},k},\widehat{g}_{k} \mid k \in \mathcal{K}\} $
and $\widehat{\mathbf{u}} \triangleq \{\widehat{u}_{\mathrm{c},k},\widehat{u}_{k} \mid k \in \mathcal{K}\} $.
Extending this approach to the power problem yields
%
\vspace{-0.04 in}
\begin{equation}
\label{Eq_Problem_WMSE_P_RS_LB}
\widehat{\mathcal{P}}_{\mathrm{RS}}(\widehat{R}_{\mathrm{t}}) \! : \!
\begin{cases}
       \underset{\widehat{R}_{\mathrm{c}},\widehat{\mathbf{c}},\mathbf{P},\widehat{\mathbf{g}},\widehat{\mathbf{u}}}
       {\min}
       \ \ \mathrm{tr}\big(\mathbf{P}\mathbf{P}^{H}\big)\\
       \text{s.t.} \\
       1 \! - \! \xi_{k}\big( \mathbf{h}_{k}, \widehat{g}_{k}, \widehat{u}_{k} \big) \! + \! \widehat{C}_{k}  \geq  \widehat{R}_{\mathrm{t}}, \forall \mathbf{h}_{k} \in \mathbb{H}_{k}, k \in  \mathcal{K} \\
        1 - \xi_{\mathrm{c},k}\big( \mathbf{h}_{k}, \widehat{g}_{\mathrm{c},k}, \widehat{u}_{\mathrm{c},k} \big)  \geq \widehat{R}_{\mathrm{c}}, \forall \mathbf{h}_{k} \in \mathbb{H}_{k},
                    k \in  \mathcal{K} \\
        \quad \sum_{k=1}^{K} \widehat{C}_{k} =  \widehat{R}_{\mathrm{c}} \\
        \quad \widehat{C}_{k} \geq 0, \; \forall k \in \mathcal{K}.
\end{cases}
\end{equation}
The semi-infiniteness is eliminated by reformulating the infinite sets of rate constraints into equivalent Linear Matrix Inequalities (LMIs) using the result in \cite{Eldar2004}, based on the $\mathcal{S}$-procedure.
The $k$th total rate constraint in \eqref{Eq_Problem_WMSE_R_RS_LB} and \eqref{Eq_Problem_WMSE_P_RS_LB} is rewritten as
\vspace{-0.08 in}
\begin{subequations}
\label{Eq_finite_WMSE_constraints_k}
\begin{align}
\widehat{u}_{k} \big( \tau_{k} +  |\widehat{g}_{k}|^{2}\sigma_{\mathrm{n}}^{2} \big) - \log_{2}(\widehat{u}_{k}) \leq \ &
1+\widehat{C}_{k} - \widehat{R}_{\mathrm{t}} \\
\label{Eq_finite_WMSE_constraints_k_2}
\left[
  \begin{array}{ccc}
    \tau_{k} - \lambda_{k} & \bm{\psi}_{k}^{H}  & \mathbf{0}^{T}  \\
     \bm{\psi}_{k}  &  \mathbf{I}  &  -\delta_{k}\mathbf{P}_{\mathrm{p}}^{H}\widehat{g}_{k}^{H} \\
     \mathbf{0}     &-\delta_{k}\widehat{g}_{k}\mathbf{P}_{\mathrm{p}} & \lambda_{k} \mathbf{I}  \\
  \end{array}
\right] & \succeq 0, \; \lambda_{k} \geq 0
\end{align}
\end{subequations}
while the $k$th common rate constraint is expressed as
\vspace{-0.08 in}
\begin{subequations}
\label{Eq_finite_WMSE_constraints_c}
\begin{align}
\widehat{u}_{\mathrm{c},k} \big(\tau_{\mathrm{c},k} +  |\widehat{g}_{\mathrm{c},k}|^{2}\sigma_{\mathrm{n}}^{2} \big) - \log_{2}(\widehat{u}_{\mathrm{c},k}) \leq \ &
1 - \widehat{R}_{\mathrm{c}} \\
\label{Eq_finite_WMSE_constraints_c_2}
\left[ \! \! \!
  \begin{array}{ccc}
    \tau_{\mathrm{c},k} - \lambda_{\mathrm{c},k} & \bm{\psi}_{\mathrm{c},k}^{H}  & \mathbf{0}^{T}  \\
     \bm{\psi}_{\mathrm{c},k}  &  \mathbf{I}  &  -\delta_{k}\mathbf{P}^{H}\widehat{g}_{\mathrm{c},k}^{H} \\
     \mathbf{0}    &-\delta_{k}\widehat{g}_{\mathrm{c},k}\mathbf{P} & \lambda_{\mathrm{c},k} \mathbf{I}  \\
  \end{array}
  \! \! \!
\right]  \!   \succeq \!    0, & \;  \lambda_{\mathrm{c},k} \! \geq \! 0
\end{align}
\end{subequations}
where $\bm{\psi}_{k}^{H} \triangleq \widehat{g}_{k}\widehat{\mathbf{h}}_{k}^{H} \mathbf{P}_{\mathrm{p}} - \mathbf{e}_{k}^{T}$
and
$\bm{\psi}_{\mathrm{c},k}^{H} \triangleq \widehat{g}_{\mathrm{c},k}\widehat{\mathbf{h}}_{k}^{H} \mathbf{P} - \mathbf{e}_{1}^{T}$.
For a detailed description of the procedure, please refer to \cite{Vucic2009a} and \cite{Tajer2011}.
Next, we develop an unified AO algorithm that solves \eqref{Eq_Problem_WMSE_R_RS_LB} and \eqref{Eq_Problem_WMSE_P_RS_LB}.
%
\subsection{Alternating Optimization Algorithm}
%
In each iteration of the algorithm, $\widehat{\mathbf{g}}$ is first optimized by solving the problems
$\underset{\widehat{g}_{\mathrm{c},k}}{\min} \underset{\mathbf{h}_{k} \in \mathbb{H}_{k}}{\max}
\varepsilon_{\mathrm{c},k}\big( \mathbf{h}_{k}, \widehat{g}_{\mathrm{c},k} \big)$
and
$\underset{\widehat{g}_{k}}{\min} \underset{\mathbf{h}_{k} \in \mathbb{H}_{k}}{\max}
\varepsilon_{k}\big( \mathbf{h}_{k}, \widehat{g}_{k} \big)$ for all $k\in \mathcal{K}$,
formulated with objective functions
$\tau_{\mathrm{c},k} +  |\widehat{g}_{\mathrm{c},k}|^{2}\sigma_{\mathrm{n}}^{2}$
and
$\tau_{k} +  |\widehat{g}_{k}|^{2}\sigma_{\mathrm{n}}^{2}$,
and constraints \eqref{Eq_finite_WMSE_constraints_c_2}  and \eqref{Eq_finite_WMSE_constraints_k_2}, respectively (weights are fixed and ignored in this step).
Such problems are posed as Semidefinite Programs (SDPs) and can be solved efficiently using interior-point methods \cite{Boyd2004}.
The resulting conservative MMSEs, $\widehat{\varepsilon}_{k}^{\mathrm{MMSE}}$ and $\widehat{\varepsilon}_{\mathrm{c},k}^{\mathrm{MMSE}}$, are used to update the weights in the next step as $\widehat{u}_{k} = 1/\widehat{\varepsilon}_{k}^{\mathrm{MMSE}}$ and $\widehat{u}_{\mathrm{c},k} = 1/\widehat{\varepsilon}_{\mathrm{c},k}^{\mathrm{MMSE}}$.
Finally, $(\mathbf{P},\widehat{\mathbf{c}})$ and the auxiliary rate variables are updated by solving a SDP formulated by fixing $(\widehat{\mathbf{g}},\widehat{\mathbf{u}})$ in \eqref{Eq_Problem_WMSE_R_RS_LB}, or \eqref{Eq_Problem_WMSE_P_RS_LB}.
This procedure is repeated in an iterative manner until convergence, which is guaranteed since the bounded objective functions behave monotonically over iterations.
However, appropriate initialization of $\mathbf{P}$ is required for the power problem to avoid feasibility issues. This is done by performing rate optimization for different power constraints in the first few iterations until a feasible solution is found, before switching to power optimization.
The conservative approximations guarantee that the AO algorithm yields feasible (although possibly sub-optimal) solutions for the original problems.
However, global optimality cannot be guaranteed even w.r.t \eqref{Eq_Problem_WMSE_R_RS_LB} and \eqref{Eq_Problem_WMSE_P_RS_LB} due to non-convexity.
Despite this sub-optimality, such algorithms were shown to perform well \cite{Tajer2011}.
\section{Simulation Results}
\label{Section_Simulation_Results}
\begin{figure}[htb]
\hspace{0.05 in}
\begin{minipage}[b]{.48\linewidth}
  \centering
  \centerline{\includegraphics[width=4.0cm]{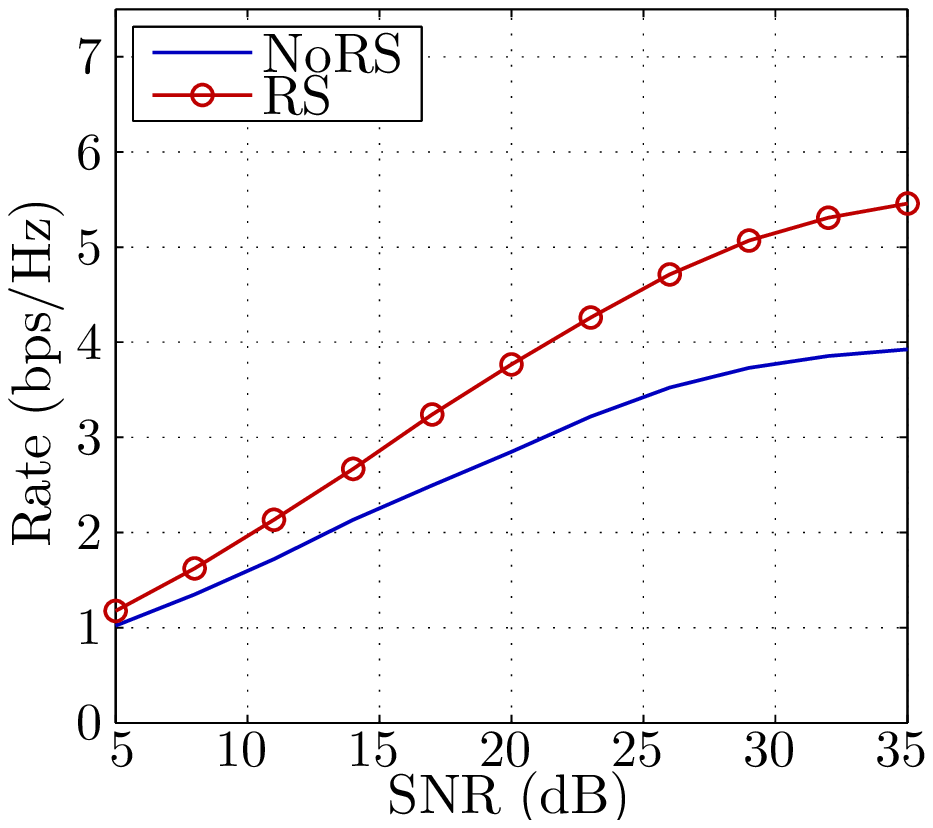}}
  \centerline{(a) $\alpha = 0$, $\delta = 0.1$}\medskip
\end{minipage}
\hspace{-0.05 in}
\begin{minipage}[b]{0.48\linewidth}
  \centering
  \centerline{\includegraphics[width=4.0cm]{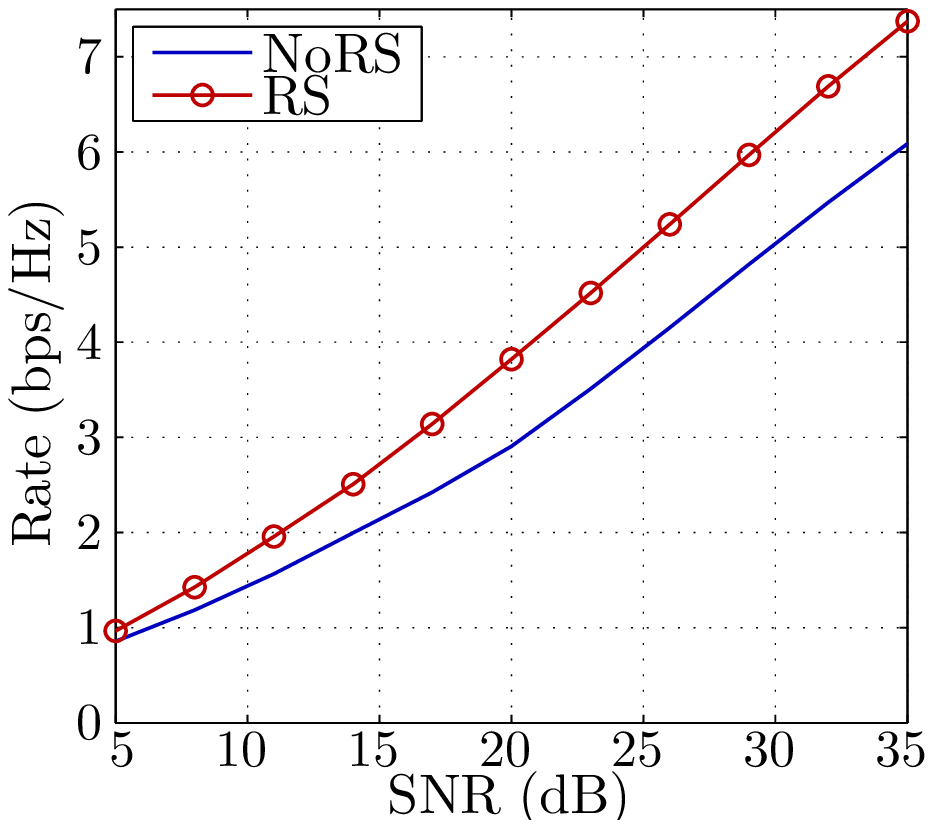}}
  \centerline{(b) $\alpha = 0.5$, $\delta = 0.1\sqrt{10P_{\mathrm{t}}^{-\alpha}}$}\medskip
\end{minipage}
\vspace{-0.15 in}
\caption{Rate performance for $K=N_{\mathrm{t}} = 3$, and $\delta_{1},\delta_{2},\delta_{3} = \delta$.}
\label{Fig_Rate}
\end{figure}
\begin{figure}[htb]
\vspace{-0.08 in}
\hspace{0.05 in}
\begin{minipage}[b]{.48\linewidth}
  \centering
  \centerline{\includegraphics[width=4.0cm]{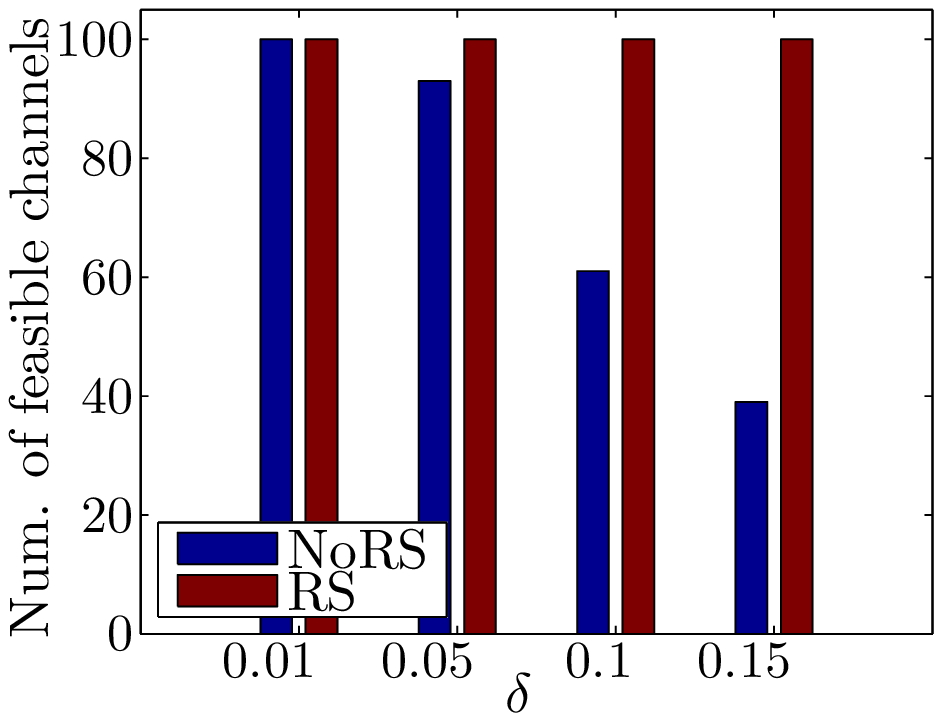}}
  \centerline{(a) Feasible channels out of 100}\medskip
\end{minipage}
\hspace{-0.05 in}
\begin{minipage}[b]{0.48\linewidth}
  \centering
  \centerline{\includegraphics[width=4.0cm]{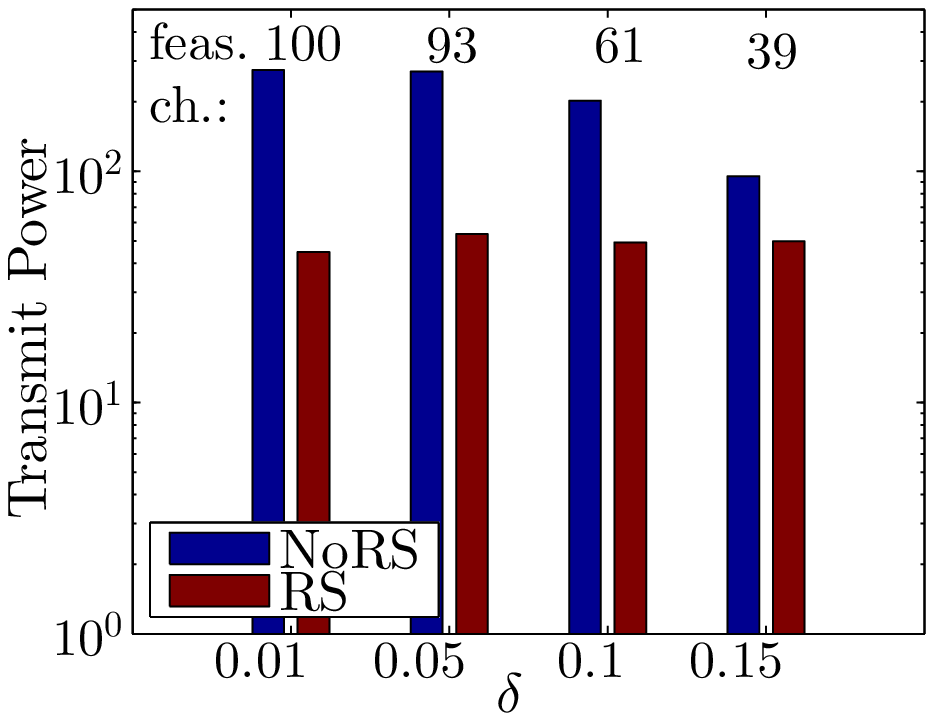}}
  \centerline{(b) Average transmit power}\medskip
\end{minipage}
\vspace{-0.15 in}
\caption{Power minimization under a rate constraint of $3.3219$ bps/Hz, for $K=N_{\mathrm{t}} = 3$, and $\delta_{1},\delta_{2},\delta_{3} = \delta$.}
\label{Fig_Power}
\vspace{-0.15 in}
\end{figure}
We consider a system with $K,N_{\mathrm{t}} = 3$, and i.i.d channel entries drawn from $\mathcal{CN}(0,1)$. We set $\sigma_{\mathrm{n}}^{2} = 1$, yielding a long-term SNR of $P_{\mathrm{t}}$.
CSIT errors are uniformly drawn from the corresponding uncertainty regions, from which channel estimates are obtained as $\widehat{\mathbf{h}}_{k} = \mathbf{h}_{k} - \widetilde{\mathbf{h}}_{k}$, $\forall k \in \mathcal{K}$.
Fig. \ref{Fig_Rate} shows the conservative worst-case rate performance averaged over $100$ channels.
The scaling uncertainty in Fig. \ref{Fig_Rate}b coincides with the non-scaling uncertainty in Fig. \ref{Fig_Rate}a at $20$ dB SNR.
It is evident that RS outperforms NoRS for all SNRs, with almost $30 \%$ improvement achieved at high SNR in Fig. \ref{Fig_Rate}a, and $25 \%$ improvement achieved at intermediate SNR in Fig. \ref{Fig_Rate}b.
At first glance, the RS rate saturation in Fig. \ref{Fig_Rate}a seems to contradict Theorem \ref{Theorem_RS_NoRS_Max_Min_DoF}.
However, this can be regarded to the design's sub-optimality and the looseness of the conservative approximation \cite{Tajer2011}, bearing in mind that Theorem \ref{Theorem_RS_NoRS_Max_Min_DoF} quantifies the optimum performance.

For power optimization, Fig. \ref{Fig_Power}a compares the feasibility of RS and NoRS over a range of $\delta$s,
where RS achieves an improvement exceeding $100 \%$ compared to NoRS at $\delta = 0.15$.
On the other hand, Fig. \ref{Fig_Power}b shows the improved power performance.
Intuitively, we expect the power gap to increase with $\delta$.
However, since infeasible channels are omitted in the average power calculation, averaging is restricted to very well conditioned channels for larger $\delta$s.
\balance
\section{Conclusion}
\label{Section_Conclusion}
In this contribution, we developed a robust RS transmission strategy to address the rate and power design problems in MU-MISO systems with CSIT uncertainties.
This builds upon the existing RS approach used to boost the sum-DoF and sum-rate performances.
We analytically proved that properly designed RS schemes achieve superior max-min DoF performances compared to their NoRS counterparts.
Moreover, we showed that RS can be used to tackle the feasibility problem appearing in NoRS power designs.
We proposed a sub-optimal unified algorithm that solves the robust RS rate and power problems based on a conservative approximation. The superior performance of the RS algorithm compared to its NoRS counterpart was demonstrated through simulations.
In the extended version of this work, we seek to develop a non-conservative robust design that achieves the theoretically anticipated non-saturating rate performance under non-scaling CSIT uncertainties.
\bibliographystyle{IEEEbib}
\bibliography{strings,References}

\begin{thebibliography}{10}

\bibitem{Vucic2009}
N.~Vucic and H.~Boche,
\newblock ``{Robust QoS-Constrained Optimization of Downlink Multiuser MISO
  Systems},''
\newblock {\em IEEE Trans. Signal Process.}, vol. 57, no. 2, pp. 714--725, Feb.
  2009.

\bibitem{Vucic2009a}
N.~Vucic, H.~Boche, and S.~Shi,
\newblock ``{Robust Transceiver Optimization in Downlink Multiuser MIMO
  Systems},''
\newblock {\em IEEE Trans. Signal Process.}, vol. 57, no. 9, pp. 3576--3587,
  Sep. 2009.

\bibitem{Payaro2007}
M.~Payar\'{o}, A.~Pascual-Iserte, and M.~A. Lagunas,
\newblock ``{Robust Power Allocation Designs for Multiuser and Multiantenna
  Downlink Communication Systems through Convex Optimization},''
\newblock {\em IEEE J. Sel. Areas Commun.}, vol. 25, no. 7, pp. 1390--1401,
  Sep. 2007.

\bibitem{Shenouda2007}
M.B. Shenouda and T.N. Davidson,
\newblock ``{Convex Conic Formulations of Robust Downlink Precoder Designs With
  Quality of Service Constraints},''
\newblock {\em IEEE J. Sel. Topics Signal Process.}, vol. 1, no. 4, pp.
  714--724, Dec. 2007.

\bibitem{Shenouda2009}
M.B. Shenouda and T.N. Davidson,
\newblock ``{Nonlinear and linear broadcasting with QoS requirements: Tractable
  approaches for bounded channel uncertainties},''
\newblock {\em IEEE Trans. Signal Process.}, vol. 57, no. 5, pp. 1936--1947,
  2009.

\bibitem{Jindal2006}
N.~Jindal,
\newblock ``{MIMO Broadcast Channels With Finite-Rate Feedback},''
\newblock {\em IEEE Trans. Inf. Theory}, vol. 52, no. 11, pp. 5045--5060, Nov.
  2006.

\bibitem{Yang2013}
S.~Yang, M.~Kobayashi, D.~Gesbert, and X.~Yi,
\newblock ``{Degrees of freedom of time correlated MISO broadcast channel with
  delayed CSIT},''
\newblock {\em IEEE Trans. Inf. Theory}, vol. 59, no. 1, pp. 315--328, 2013.

\bibitem{Hao2013}
C.~Hao and B.~Clerckx,
\newblock ``{MISO BC with imperfect and (Un)matched CSIT in the frequency
  domain: DoF region and transmission strategies},''
\newblock in {\em Proc. IEEE PIMRC 2013}, Sep. 2013, pp. 1--6.

\bibitem{Hao2015}
C.~Hao, Y.~Wu, and B.~Clerckx,
\newblock ``{Rate Analysis of Two-Receiver MISO Broadcast Channel With Finite
  Rate Feedback: A Rate-Splitting Approach},''
\newblock {\em IEEE Trans. Commun.}, vol. 63, no. 9, pp. 3232--3246, Sept 2015.

\bibitem{Joudeh2015}
H.~Joudeh and B.~Clerckx,
\newblock ``Sum rate maximization for {MU-MISO} with partial {CSIT} using joint
  multicasting and broadcasting,''
\newblock in {\em Proc. IEEE ICC 2015}, Jun. 2015, pp. 6349--6354.

\bibitem{Dai2015a}
M.~Dai, B.~Clerckx, D.~Gesbert, and G.~Caire,
\newblock ``{A Rate Splitting Strategy for Massive MIMO with Imperfect CSIT},''
\newblock {\em submitted to IEEE Trans. Wireless Commun.}, available at
  http://arxiv.org/abs/1512.07221.

\bibitem{Joudeh2015a}
H.~Joudeh and B.~Clerckx,
\newblock ``Achieving {Max-Min} fairness for {MU-MISO} with partial {CSIT:} a
  multicast assisted transmission,''
\newblock in {\em Proc. IEEE ICC 2015}, Jun. 2015, pp. 6355--6360.

\bibitem{Caire2010}
G.~Caire, N.~Jindal, M.~Kobayashi, and N.~Ravindran,
\newblock ``{Multiuser MIMO Achievable Rates With Downlink Training and Channel
  State Feedback},''
\newblock {\em IEEE Trans. Inf. Theory}, vol. 56, no. 6, pp. 2845--2866, Jun.
  2010.

\bibitem{Tajer2011}
A.~Tajer, N.~Prasad, and X.~Wang,
\newblock ``{Robust Linear Precoder Design for Multi-Cell Downlink
  Transmission},''
\newblock {\em IEEE Trans. Signal Process.}, vol. 59, no. 1, pp. 235--251, Jan.
  2011.

\bibitem{Wiesel2006}
A.~Wiesel, Y.C. Eldar, and S.~Shamai,
\newblock ``{Linear precoding via conic optimization for fixed MIMO
  receivers},''
\newblock {\em IEEE Trans. Signal Process.}, vol. 54, no. 1, pp. 161--176, Jan.
  2006.

\bibitem{Sidiropoulos2006}
N.D. Sidiropoulos, T.N. Davidson, and Z.-Q. Luo,
\newblock ``Transmit beamforming for physical-layer multicasting,''
\newblock {\em IEEE Trans. Signal Process.}, vol. 54, no. 6, pp. 2239--2251,
  Jun. 2006.

\bibitem{Christensen2008}
S.S. Christensen, R.~Agarwal, E.~Carvalho, and J.M. Cioffi,
\newblock ``{Weighted sum-rate maximization using weighted MMSE for MIMO-BC
  beamforming design},''
\newblock {\em IEEE Trans. Wireless Commun.}, vol. 7, no. 12, pp. 4792--4799,
  Dec. 2008.

\bibitem{Shi2011}
Q.~Shi, M.~Razaviyayn, Z.-Q. Luo, and C.~He,
\newblock ``{An Iteratively Weighted MMSE Approach to Distributed Sum-Utility
  Maximization for a MIMO Interfering Broadcast Channel},''
\newblock {\em IEEE Trans. Signal Process.}, vol. 59, no. 9, pp. 4331--4340,
  Sept 2011.

\bibitem{Razaviyayn2013b}
M.~Razaviyayn, M.~Hong, and Z.-Q. Luo,
\newblock ``{Linear transceiver design for a MIMO interfering broadcast channel
  achieving max--min fairness},''
\newblock {\em Signal Processing}, vol. 93, no. 12, pp. 3327 -- 3340, 2013.

\bibitem{Eldar2004}
Y.C. Eldar and N.~Merhav,
\newblock ``A competitive minimax approach to robust estimation of random
  parameters,''
\newblock {\em IEEE Trans. Signal Process.}, vol. 52, no. 7, pp. 1931--1946,
  Jul. 2004.

\bibitem{Boyd2004}
S.~P. Boyd and L.~Vandenberghe,
\newblock {\em {Convex Optimization}},
\newblock Cambridge university press, 2004.

\end{thebibliography}
\balance

\end{document}